\documentclass[reprint,showpacs,amsmath,amssymb,aps,prfluids]{revtex4}
\usepackage{amsmath}
\usepackage{float}
\usepackage{graphicx}
\usepackage{color,xcolor}
\usepackage[amssymb]{SIunits}
\usepackage[colorlinks,citecolor = blue, linkcolor=red,hyperindex,CJKbookmarks]{hyperref}

\begin{document}

\title{Circular objects do not melt the slowest in water} 
\author{Rui Yang$^{1}$}
\author{Thijs van den Ham$^{1}$}
\author{Roberto Verzicco$^{1,2,3}$}
\author{Detlef Lohse$^{1,4}$}
\author{Sander G. Huisman$^{1}$}

\affiliation{$^1$Physics of Fluids Group and Max Planck Center for Complex Fluid Dynamics, J. M. Burgers Centre for Fluid Dynamics, University of Twente, P.O. Box 217, 7500AE Enschede, The Netherlands\\
$^2$Dipartimento di Ingegneria Industriale, University of Rome `Tor Vergata', Roma 00133, Italy\\
$^3$Gran Sasso Science Institute, Viale F. Crispi, 7 67100 L'Aquila, Italy\\
$^4$Max Planck Institute for Dynamics and Self-Organization, Am Fa\ss berg 17, 37077 G\"{o}ttingen, Germany}

\date{\today}

\begin{abstract}
We report on the melting dynamics of ice suspended in fresh water and subject to natural convective flows. Using direct numerical simulations we investigate the melt rate of ellipsoidal objects for $2.32\times 10^4 \leq \text{Ra} \leq 7.61\times 10^8$, where \text{Ra} is the Rayleigh number defined with the temperature difference between the ice and the surrounding water. We reveal that the system exhibits non-monotonic behavior in three control parameters. As a function of the aspect ratio of the ellipsoidal, the melting time shows a distinct minimum that is different from a disk which has the minimum perimeter. Furthermore, also with \text{Ra} the system shows a non-monotonic trend, since for large \text{Ra} and large aspect ratio the flow separates, leading to distinctly different dynamics. Lastly, since the density of water is non-monotonic with temperature, the melt rate depends non-monotonically also on the ambient temperature, as for intermediate temperatures ($\unit{4}{\celsius}$--$\unit{7}{\celsius}$) the flow is (partially) reversed. In general, the shape which melts the slowest is quite distinct from that of a disk.
\end{abstract}

\maketitle
Melting plays a significant role in various natural phenomena, including oceanography, geophysics, and 
astrophysics, and in engineering applications like in process technology and thermal energy storage. As an example of the classical Stefan problem \cite{rubinstein1971stefan}, melting phenomena can exhibit quite some 
 complexities \cite{cenedese2022icebergs,weady2022anomalous}, particularly in their
  interaction with buoyant flows. These complexities  are
   evident in scenarios involving intricate multiscale geological morphologies and accelerated melt rates due to ambient flows. Understanding these phenomena and the underlying physical mechanisms is crucial, especially given the escalating melt rates observed in global ice reserves \cite{chen2006satellite}. In this paper we 
   study the complexities of the melting processes in idealized canonical geometries to better understand the underlying physics.

Despite extensive research on ice melt rates using models \cite{weeks1973icebergs,fitzmaurice2017nonlinear,martin2010parameterizing}, experiments \cite{hao2001melting,hao2002heat,weady2022anomalous,waasdorp2023melting}, and simulations \cite{esfahani2018basal,favier2019,yang2023morphology,yang2023icemelting}, the potential effects of geometry on the melt rate are generally ignored. Not only icebergs and ice floes exhibit considerable shape and size variations \cite{gherardi2015characterizing,hewitt2020subglacial}, with sizes ranging from a few meters to several hundred kilometers, but also in dissolution processes the size spans several orders of magnitude. In lab experiments in which ice was melted by external flows the overall melt rate has been found to depend on the aspect ratio \cite{hester2021aspect,cenedese2022icebergs}. Despite its importance for the ice melt rate, the effect of shape is still poorly understood. Therefore, to deepen our understanding and improve the iceberg melting predictions, it is imperative to consider aspect ratio and shape in models of iceberg melt rates. The natural fundamental question we address here are: how does the melt rate depend on the shape, and what is the optimal shape (defined as the slowest-melting shape for fixed initial volume)?

To answer these questions consider the three mechanisms of heat transfer: conduction, radiation, and convection. For conduction and radiation the transfer is proportional to the area and we can therefore restate the problem as finding a shape with minimal surface area. From the isoperimetric inequality \cite{osserman1978isoperimetric} we know that an $n$-ball---with its $\mathrm{O}(n)$ symmetry---gives us the shape for which the perimeter (a circle in 2D or a sphere in 3D) is minimized. However, for convection the gravitational acceleration breaks this symmetry and therefore an $n$-ball is no longer the shape that melts the slowest. This means that, whereas for conduction and radiation an initial ball shape stays \textit{self-similar} during the melting process, for convection the shape evolves over time and is \textit{not} self-similar. Previous studies also highlight the complex interplay between ice melting and ambient flows, leading to distinct morphologies and melt rates \cite{ristroph2012sculpting,hao2002heat,weady2022anomalous,yang2023morphology}.

\begin{figure}
 \includegraphics[width=\columnwidth]{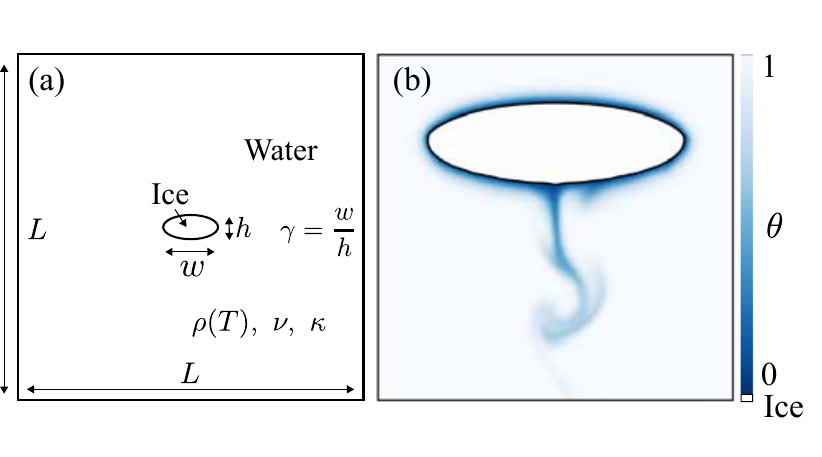}
 \caption{(a) Simulation setup. The initial effective diameter of ice is fixed as $D=\sqrt{hw}=0.1L$. (b) Zoom-in view of the temperature field of a melting ice block with an aspect ratio $\gamma=3$. $\theta$ is the dimensionless temperature.}
 \label{fig:fig1}
\end{figure}

In this Letter, the impact of the initial ice shape on the melt rates are investigated by means of direct numerical simulations and theoretical analysis. We focus on the scenario of ice melting in a water-filled box, where natural convection due to buoyancy dominates the flow. Our aim is to understand how the aspect ratio affects the ice melt rates, with the help of direct numerical simulations. We numerically integrate the Navier--Stokes and advection-diffusion equations to find the evolution of the velocity field $\boldsymbol{u}(\boldsymbol{x},t)$ and the temperature field $T(\boldsymbol{x},t)$, respectively. Note that for melting of ice in cold water around $\unit{4}{\celsius}$, the density anomaly of water is key \cite{wang2021ice,wang2021growth,wang2021equilibrium,yang2022abrupt}; we approximate it as:
\begin{align}
 \rho=\rho_0(1-\beta|T-T_{\text{max}}|^q),
\label{eq:density}
\end{align}
where $\beta=9.3\times10^{-6}\ (^\circ C)^{-q}$ is the generalized thermal expansion coefficient, with the exponent $q=1.895$ \cite{gebhart1977new} and $T_{\text{max}}=\unit{4}{\celsius}$ for fresh water. The melting process is modeled by the phase-field method, which has been widely used and verified in previous studies \cite{favier2019,hester2020improved,hester2021aspect,couston2021topography,yang2022abrupt,yang2023morphology}. In this technique, the phase field variable $\phi$ is integrated in space and time and smoothly transitions from a value of $1$ in the solid phase to a value of $0$ in the liquid phase. In the simulations, we prescribe an ice object with an initial cross-sectional area $A_0$ and effective diameter $D=2\sqrt{A_0/\pi}$ at the center of a square domain with the side length $L=10D$ (corresponding area ratio $A_0/L^2=0.8\%$), see Fig.\ \ref{fig:fig1}(a). Two-dimensional and three-dimensional simulations (with size $L$ in the third dimension) are performed. All boundaries are adiabatic with free-slip conditions on the velocity field. An example of the temperature field of a melting ice object is shown in Fig.\ \ref{fig:fig1}(b). More details on the governing equations and the numerical method can be found in the Supplementary Materials.

The control parameters of the system are the Rayleigh number $\text{Ra}$ (dimensionless buoyancy strength), the aspect ratio $\gamma$ of the initial ice shape which is defined as the ratio of its width $w$ and height $h$, the Prandtl number $\text{Pr}$, which is the ratio between the kinematic viscosity $\nu$ and the thermal diffusivity $\kappa$, and the Stefan number $\text{Ste}$, which is the ratio between the latent heat $\mathcal{L}$ and the sensible heat:
\begin{align}
\text{Ra}&=\frac{g \beta \Delta^q D^3}{\nu\kappa},& \gamma&=\frac{w}{h},& \text{Pr}&=\frac{\nu}{\kappa},& \text{Ste}&=\frac{\mathcal{L}}{c_p\Delta}.
\end{align}
Here, $g$ is the gravitaional acceleration, $c_p$ is the specific heat capacity, and $\Delta$ is the temperature difference between the ambient water and the ice. Given the large parameter space, some of the control parameters have to be fixed in order to make the study feasible. We fix $\text{Pr}=7$ and $\text{Ste}=4$ as the values for water at $\unit{20}{\celsius}$. Our simulations cover a parameter range of $10^4 \leq \text{Ra} \leq 10^9$ (corresponding to ice diameters $\unit{5}{\milli\meter} \leq D \leq \unit{160}{\milli\meter}$) and $0.5 \leq \gamma \leq 4$.

\begin{figure}
 \includegraphics[width=0.9\columnwidth]{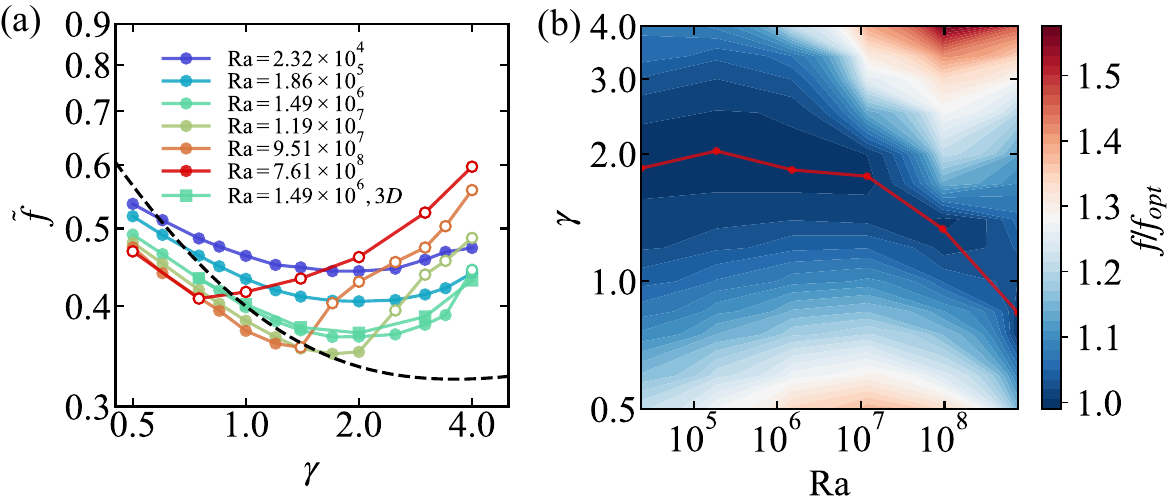}
 \caption{(a) Rescaled melt rate $\tilde f = f\text{Ra}^{1/4}$ as a function of the aspect ratio $\gamma$ for different $\text{Ra}$. The dashed line shows the theoretical prediction for the curve of the minima from eq.~(\ref{eq:tfinal}). (b) The heat map of the rescaled melt rate (by the optimal melt rate $f_{\text{opt}}$) as a function of $\text{Ra}$ and $\gamma$. The red line shows the optimal aspect ratio $\gamma_{\text{min}}$ with the minimum melt rate, which is obtained by a quadratic fitting of the data points around the minimum.}
 \label{fig:fig2}
\end{figure}

We first investigate dependences of the melt rate dependence on the aspect ratio $\gamma$ and $\text{Ra}$, as shown in Fig.~\ref{fig:fig2}(a). The time it takes for the ice to completely melt is $t_f$, and we define $f=1/t_f$ as mean melt rate. The trend of rescaled melt rate $\tilde{f}=fRa^{1/4}$ depends not only on $\gamma$ but also on $\text{Ra}$. Here we rescale the melt rate to make the trends for different \text{Ra} comparable. For increasing $\gamma$, the melt curves exhibit a non-monotonic trend, with the rescaled melt rate first decreasing and then increasing. As explained above, in the absence of gravity ($\text{Ra}=0$), the minimum melt rate would occur at an aspect ratio $\gamma_{\text{min}}=1$ (disk shape) as that minimizes the perimeter. As $\text{Ra}$ increases, $\gamma_{\text{min}}$ first increases to around $\gamma_{\text{min}}\approx2$. The same trend is also observed for 3D simulations of elliptical cylinders with cross-sections of different aspect ratios. However, as $\text{Ra}$ increases further ($\text{Ra}\ge3\times10^8$), remarkably $\gamma_{\text{min}}$ starts to decrease even below $1$. The same observation can also be seen from the heatmap of the normalized melt rate in Fig.\ \ref{fig:fig2}(b).

Our findings demonstrate that in absence of external flow, ice with an elliptical shape (when the short axis is aligned with gravity) can melt up to 10\% slower than a disk shape. This previously neglected shape factor may have implications for accurately estimating the melt rate of icebergs in previous models \cite{weeks1973icebergs,martin2010parameterizing,fitzmaurice2017nonlinear} that neglected it. The counter-intuitive nature of this phenomenon demands an explanation and raises further questions about its physical mechanisms.

\begin{figure}
 \includegraphics[width=1\columnwidth]{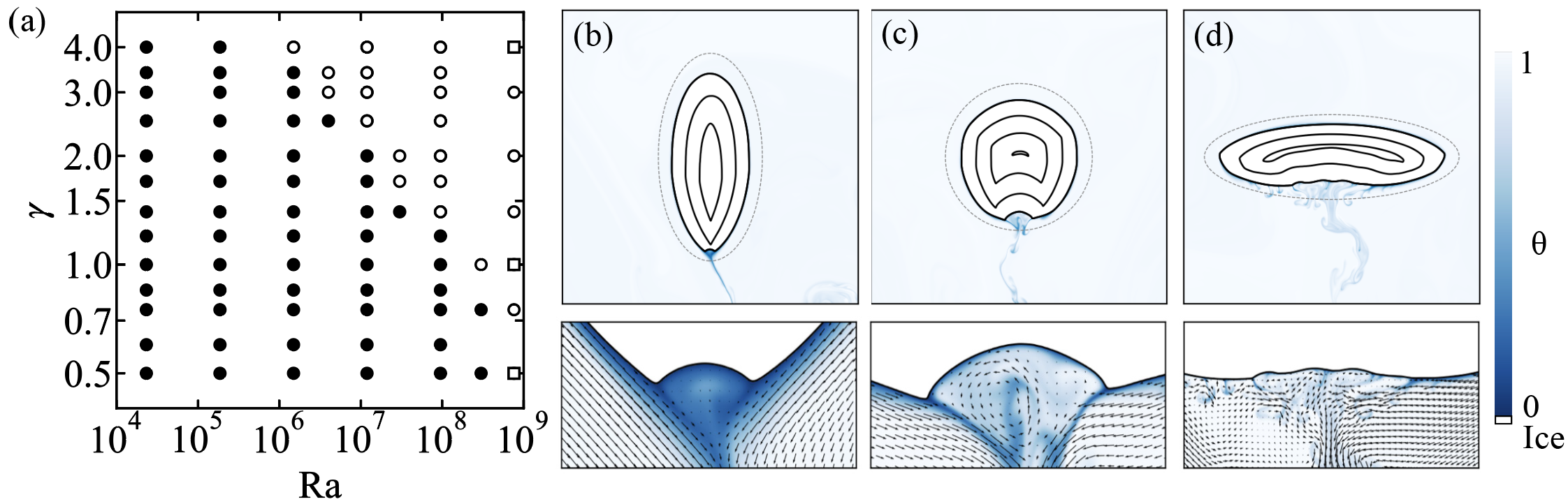}
 \caption{((a) The $\text{Ra}$--$\gamma$ phase diagram of cavity formation. The black disks mean no-cavity formation during melting and the circles mean that cavity formation is observed. The squares in (a) correspond to the cases shown in (b--d) where the instantaneous temperature field and a zoom-in of the cavity at the bottom for $\text{Ra}=7.61\times10^8$ and (b) $\gamma=0.5$, (c) $1$, and (d) $3$. The thin dashed gray lines show the initial shape of the object, and future shapes are indicated with solid lines inside the objects.}
 \label{fig:fig3}
\end{figure}

To quantitatively explain the result, we consider the total ice mass budget for the melting process:
\begin{align}
    \frac{dA(t)}{dt}=P(\gamma,t)v_n, \label{eq:dAdt}
\end{align}
where $A$ is the total area of ice (in 2D), $P(\gamma,t)$ is the perimeter as a function of aspect ratio and time, and $v_n$ is the surface-averaged melt velocity. Assuming the elliptical shape does not significantly change with time, we have for the perimeter of an ellipse:
\begin{align}
   P(\gamma,t) = 4\sqrt{\frac{A(t)}{\pi\gamma}}E\left(1-\gamma ^2\right), \label{eq:P}
\end{align}
where $E$ is the complete elliptic integral of the second kind. The presence of a flow generated from natural convection increases the temperature gradient at the ice front and makes $v_n$ non-uniform. To estimate $v_n$, we consider the Stefan boundary condition, i.e.\ that the dimensionless surface-averaged melt velocity $\tilde{v}_n$ is related to the surface-averaged heat flux $\overline{\text{Nu}}$:
\begin{align}
   \tilde{v}_n=\frac{v_n}{U_0}=-\frac{1}{U_0}\frac{\kappa c_p}{\mathcal{L}}\frac{\partial T}{\partial n}=\frac{\overline{\text{Nu}}}{\text{Ra}^{1/2}\text{Pr}^{1/2}A^{1/2}\text{Ste}}, \label{eq:vn}
\end{align}
where $\overline{\text{Nu}}$ can be estimated from the scaling of $\text{Nu}$ for a laminar boundary layer \cite{bejan1993} by:
\begin{align}
   \overline{\text{Nu}}\propto \text{Ra}_{h}^{1/4} \label{eq:Numean}
\end{align}
with $\text{Ra}_{h}=\text{Ra}(h/D)^3$ the Rayleigh number defined by the cross-sectional height $h$ in the gravity direction. By substituting Eqs.\ (\ref{eq:P}), (\ref{eq:vn}), and (\ref{eq:Numean}) into Eq.\ (\ref{eq:dAdt}), we obtain:
\begin{align}
    f\propto P(\gamma)\gamma^{-3/8}, \label{eq:tfinal}
\end{align}
which provides the overall melt rate dependence on $\gamma$. The factor $\gamma^{-3/8}$ originates from the scaling of $\overline{\text{Nu}}$, representing the effect of convective flow on melt rate, and
 $P(\gamma)$ represents the effect of the perimeter size on the melt rate with the minimum at $\gamma=1$, which dominates when the ambient flow is absent or weak. $\gamma^{-3/8}$ displays a monotonic decrease in melt rate. This can be physically understood by the natural convective flows induced by melting. The effective length scale for natural convection is
 the vertical length, which is larger for smaller $\gamma$, resulting in larger 
  $Ra_h$ and therefore stronger convection, i.e., 
   higher melt rates. The total melt rate, the product of $P(\gamma)$ and $\gamma^{-3/8}$, features a non-monotonic trend with a minimum at $\gamma>1$, which is shown in fig.~\ref{fig:fig2}(a). Although the curve differs from our results since the actual melting dynamics is far more complicated (e.g.\ $\gamma$ is changing with time, see Supplementary Material, and Eq.\ \ref{eq:Numean} does not take into consideration any morphology changes), the trend agrees with our explanation for $\gamma_{\text{min}}>1$.

However, as \text{Ra} increases further, $\gamma_{\text{min}}$ decreases even below $1$, which is opposite to the previous prediction. This behavior of $\gamma_{\text{min}}$ is due to a transition associated with cavity formation at the bottom of the melt shape, as seen in Figs.~\ref{fig:fig3}(b--d). The parameter space in the $(\gamma,Ra)$ place is shown in figs.~\ref{fig:fig3}(a) and features a regime with transition from no-cavity to cavity formation and a regime without this transition. Figs.\ \ref{fig:fig3}(b--d) shows the instantaneous flow patterns and cavity structures for varying $\gamma$ values. This cavity formation is due to the onset of flow separation, a phenomenon that is commonly observed in flows around (heated) objects \cite{bejan1993}. The melt water flows along the surface at the top and detaches before reaching the bottom. The occurrence of flow separation leads to enhanced mixing, triggering an increased local heat flux. This enhanced local melt rate results in the development of a cavity at the bottom of the ice object. This transition can be seen for either high values of $\gamma$ or high values of $\text{Ra}$.

Moreover, the regime transition associated with cavity formation aligns with the corresponding trend transition in the melt rate, as showcased in Fig.\ \ref{fig:fig2}(a). Circles (disks) in the graph correspond to cases where cavities manifest (no cavities appear). This is consistent because the enhanced local melt rate significantly influences the overall melt rate $f$, leading to a significant increase in $f$ in these cases. This effect is particularly pronounced for larger $\gamma$ values, and it explains why $\gamma_{\text{min}}$ shifts towards smaller values, revealing the complexity in the relationship between aspect ratio and melt rates.

\begin{figure}
 \includegraphics[width=\columnwidth]{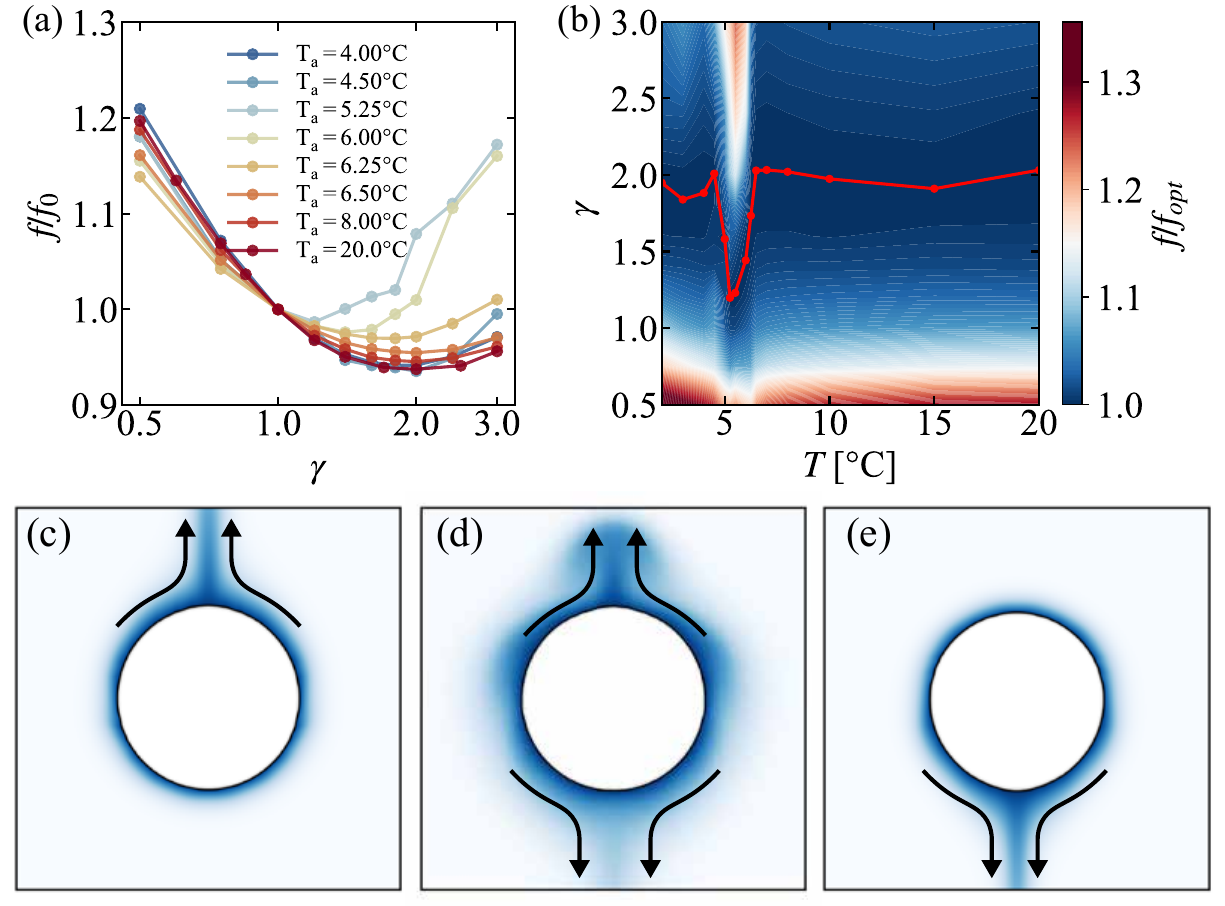}
 \caption{(a) Rescaled melt rate as a function of the aspect ratio $\gamma$ for different $T$ with a fixed $\text{Ra}=1.86\times10^5$. The melt rates are rescaled by the melt rate of the disk ($\gamma=1$) at the same ambient temperature. (b) The heat map of the rescaled melt rate as a function of $T$ and $\gamma$. The red line shows the optimal aspect ratio $\gamma_{\text{min}}$ with the minimum melt rate, which is obtained by a quadratic fitting of the data points from simulation results. (c--d) The snapshot of temperature fields at the ambient temperatures $T=\unit{4}{\celsius}$, $\unit{5.25}{\celsius}$, and $\unit{8}{\celsius}$, with the same color bar as in Fig.\ \ref{fig:fig3}. The arrows show the direction of the main flow.}
 \label{fig:fig4}
\end{figure}

Until now we only considered the ambient temperature $T_a$ as $\unit{20}{\celsius}$. However, the density of water follows a non-monotonic relationship with temperature, see Eq.\ (\ref{eq:density}). This effect can cause distinct flow regimes and ice melting morphologies for different ambient temperatures \cite{wang2021growth,wang2021equilibrium,wang2021ice}. We therefore investigate the effect of the ambient temperature on $f$ and $\gamma_{\text{min}}$ for fixed $\text{Ra}=1.86\times10^5$. Fig.\ \ref{fig:fig4}(a) shows the dependence of $\gamma$ on the overall melt rate at different ambient temperatures. One can see similar trends and values of $\gamma_{\text{min}}$ for large ambient temperature ($T_a \geq \unit{8}{\celsius}$) and small ambient temperature ($T_a \leq \unit{5}{\celsius}$). However, $\gamma_{\text{min}}$ decreases towards $1$ for intermediate temperatures ($\unit{5}{\celsius} \leq T_a \leq \unit{8}{\celsius}$). The same trend can also be seen from the heat map of the normalized melt rate in Fig.\ \ref{fig:fig4}(b), where a sharp transition of $\gamma_{\text{min}}$ (the red line) towards disk shape ($\gamma=1$) at intermediate temperatures exists.

The transition at play can be understood by the density anomaly effect, which results in distinctive flow structures for different ambient temperatures, as elucidated in Figs.\ \ref{fig:fig4}(c--e). For low ambient temperatures (Fig.\ \ref{fig:fig4}(c)), the melt water simply ascends due to its lower density at $\unit{0}{\celsius}$ as compared to the density of the surrounding water. For high ambient temperature (Fig.\ \ref{fig:fig4}(e)), the melt water's density remains higher than that of the ambient water, which means only downward flow motion. However, as the temperature of the ambient water is slightly above $\unit{4}{\celsius}$ (Fig.\ \ref{fig:fig4}(d)), the melt water adopts a bidirectional flow pattern, having both upward and downward motion, which Weady et al.\ also observe for a melting cylinder \cite{weady2022anomalous}. Initially, the melt water ($\unit{0}{\celsius}$) moves upwards. As the ambient water mixes with and warms up the melt water, the density of the melt water increases, surpassing that of the surrounding water, causing it to then flow downwards. 

 More details of the flow dynamics under different ambient temperatures can be appreciated in the movies of the Supplementary Materials. Around the intermediate temperatures, where both upward and downward flows occur, the cold melt water stays closer to the object for a longer time and also leads to a more uniform flow around the ice. Consequently, the asymmetry induced by convection weakens, and the length of the perimeter becomes dominant in determining the melt rate, which causes a transition of the optimal shape close to a disk ($\gamma_{\text{min}} \approx 1$).

In summary, we conducted comprehensive numerical investigations on elliptical shapes across various aspect ratios ($\gamma$), Rayleigh numbers $\text{Ra}$, and ambient temperatures $T_a$. Our primary focus was on understanding the behavior of the overall melt rate $f$ and the corresponding shape evolution. Notably, our findings revealed a 
non-monotonic dependence of the melt rate $f$ on these control parameters $\text{Ra}$, $\gamma$, and $T_a$. $f$ was found to be non-monotonic with $\gamma$ such that for low $\text{Ra}$ elliptically-shaped ice with $\gamma\approx2$ melts $\approx 10\%$ slower than disk-shaped ice. Moreover $\gamma_{\text{min}}$ initially increases and then decreases with rising Rayleigh number $\text{Ra}$, which we were able to explain using physical rationale. We identified a direct correlation between the reduction of $\gamma_{\text{min}}$ and the emergence of cavity formations due to flow separation as $\text{Ra}$ increases. This phenomenon significantly amplifies the local melt rate, particularly for larger $\gamma$ values. Our investigation was further extended to explore the interplay of $\gamma_{\text{min}}$ with the ambient temperature $T_a$, where we uncovered another non-monotonic relationship---essential for accurate predictions of melting in fresh water. This behavior stems from the density anomaly of water around $\unit{4}{\celsius}$, causing alterations in the flow dynamics around the ice and consequently influencing the melt rate.

Overall, our study offers comprehensive insight into the intricate relationship between the shape-dependent ice melt rate and the gravitational symmetry breaking by convection. These findings have significant relevance to the modeling of large-scale geophysical and climatological or industrial processes. 

Our physical explanation of the shift of $\gamma_{\text{min}}$ can also be extended to other moving boundary problems, such as erosion \cite{ristroph2012}, ablation, and dissolution \cite{wykes2018,mac2015}. Future studies may investigate the ice melting dynamics in salty water, where double-diffusive convection plays an important role and can result in intriguing flow patterns \cite{huppert1978melting,yang2023icemelting}.

\begin{acknowledgments}
We acknowledge the European Union (ERC, MeltDyn, 101040254) for funding this study, PRACE for awarding us access to MareNostrum in Spain at the Barcelona Computing Center (BSC) under the projects 2020235589 and 2021250115, and the German Science Foundation (DFG) through the Priority Programme SPP 1881 ``Turbulent Superstructures'' for funding. We also thank Sissi de Beer for the discussions.
\end{acknowledgments}

\section*{Methods} 

We consider an incompressible flow non-dimensionalized by the free-fall velocity $U_f=\sqrt{\beta g\Delta^q D}$ as the velocity scale, the ice effective diameter $D$ as the length scale, and the temperature difference between the ambient water and the ice $\Delta$ as the temperature scale. The non-dimensionalized quantities include (two) three velocity components $u_i$ with ($i=1,2$) $i=1,2,3$ for our simulations in (2D) 3D, the pressure $p$, the temperature $\theta$, and the phase field scalar $\phi$. The dimensionless governing equations read:
\begin{align*}
\boldsymbol{\nabla} \cdot \boldsymbol{u}&=0, \\
\frac{\partial}{\partial t} \boldsymbol{u}+(\boldsymbol{u} \cdot\boldsymbol{\nabla}) \boldsymbol{u}&=-\boldsymbol{\nabla} p+\sqrt{\frac{\text{Pr}}{\text{Ra}}}\left(\boldsymbol{\nabla}^2 \boldsymbol{u}-\frac{\phi \boldsymbol{u}}{\eta}\right)+|\theta-\theta_m|^q \boldsymbol{e}_z, \\
\frac{\partial}{\partial t} \theta+(\boldsymbol{u} \cdot\boldsymbol{\nabla})\theta&=\frac{1}{\sqrt{\text{Ra}\text{Pr}}} \boldsymbol{\nabla}^2 \theta+ \text{Ste} \frac{\partial}{\partial t} \phi, \\ 
\frac{\partial \phi}{\partial t}&=\frac{6}{5  C\text{Ste}{\sqrt{\text{Ra}\text{Pr}}}}\left[\boldsymbol{\nabla}^2 \phi-\frac{1}{\varepsilon^2} \phi(1-\phi)(1-2 \phi+C \theta)\right],
\end{align*}
where $\theta_m=T_c/\Delta$ is the nondimensionlized maximum temperature with $T_c=\unit{4}{\celsius}$, $\varepsilon$ is the diffusive interface thickness, which is typically set to be the mean grid spacing. $C$ is the phase mobility parameters related to the Gibbs--Thompson relation. We choose $C = 10$ and avoid extreme values of $\gamma$, where the high curvature regions might be inaccurate. More details can be found in previous studies \cite{yang2023morphology,hester2020improved}. Simulations are performed using the second-order staggered finite difference code AFiD, which has been extensively validated and used to study a wide range of turbulent flow problems \cite{ostilla2015multiple,yang2016pnas,liu2022heat}, including phase-change problems \cite{yang2022abrupt,yang2023morphology}. The phase field method is applied to model the phase-change process, which has been widely used in previous studies \cite{favier2019,hester2021aspect,couston2021topography,ravichandran2021melting,ravichandran_combined_2022,yang2023morphology}. The flow is confined to a square box with all no-slip boundary conditions for the velocity and is adiabatic for the temperature. A multi-resolution method \cite{ostilla2015multiple} is applied to the phase field.

\section*{Convergence test}

\begin{figure}
\centering
  \includegraphics[width=0.8\columnwidth]{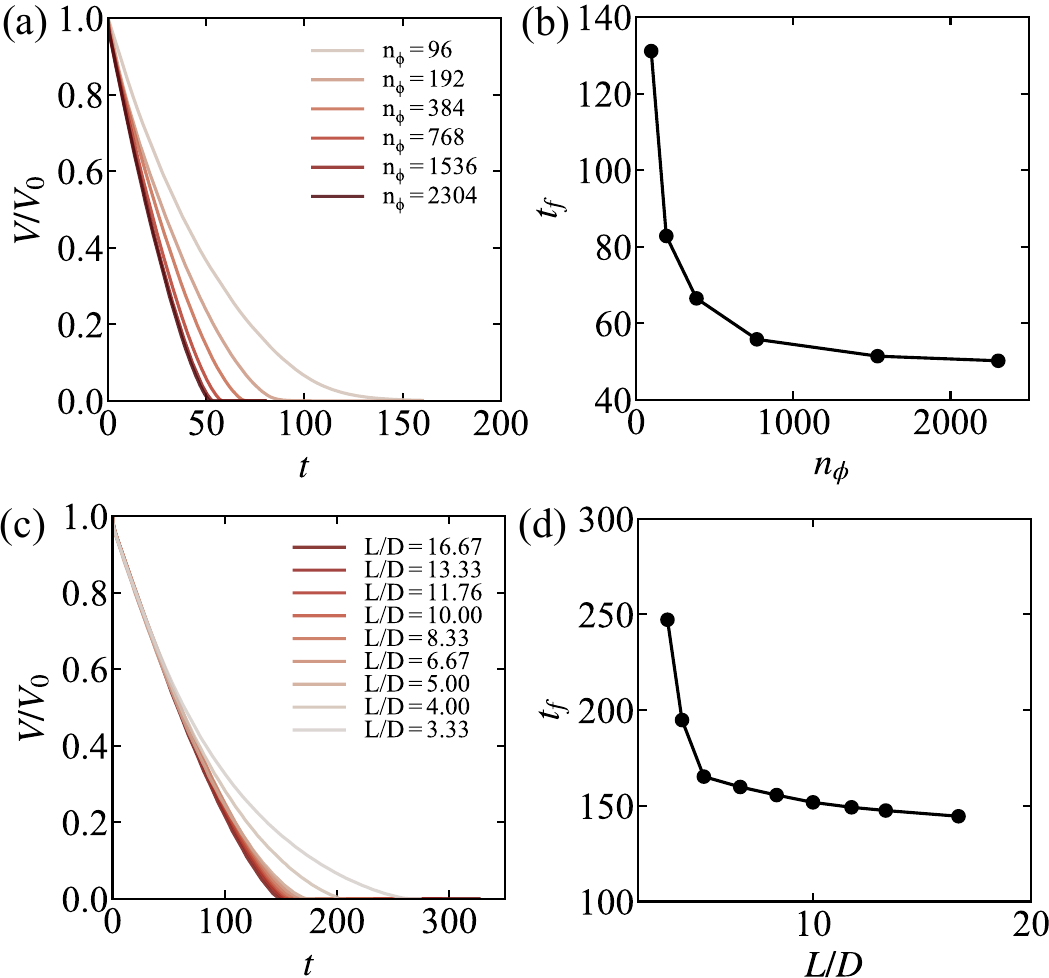}
  \caption{(a) Normalized ice volume as a function of time with different refined resolutions at $\text{Ra}=1.85\times10^5$ and $\gamma=1$. (b) The corresponding total melt rate $t_f$ (from beginning to completely melted) as a function of resolution. Our final choice of resolution for this case is 1536. (c) Normalized ice volume as a function of time with different domain-diameter ratios $(L/D)$ at $\gamma=1$. (d) The corresponding total melt rate $t_f$ as a function of $L/D$. Our final choice of resolution for this case is $L/D=10$. }
  \label{fig:suppfig1}
\end{figure}

The resolution convergence test is shown in figs.~\ref{fig:suppfig1}(a--b) for $\text{Ra}=1.85\times10^5$. Here we applied a refined ($3\times$) grid for the phase field, which shows convergence at $n_\phi=1536$. Note that the finite size of the computation domain could feasibly affect the ice melt rate, as the melt water also accumulates at the bottom. To ensure that this does not affect our results much, we performed a test for disks shaped with different values of $L/D$, see figs.~\ref{fig:suppfig1}(c--d), which produced a converging melt rate at $L/D=10$.

\section*{Nu(Ra) scaling}
\begin{figure}
\centering
  \includegraphics[width=0.8\columnwidth]{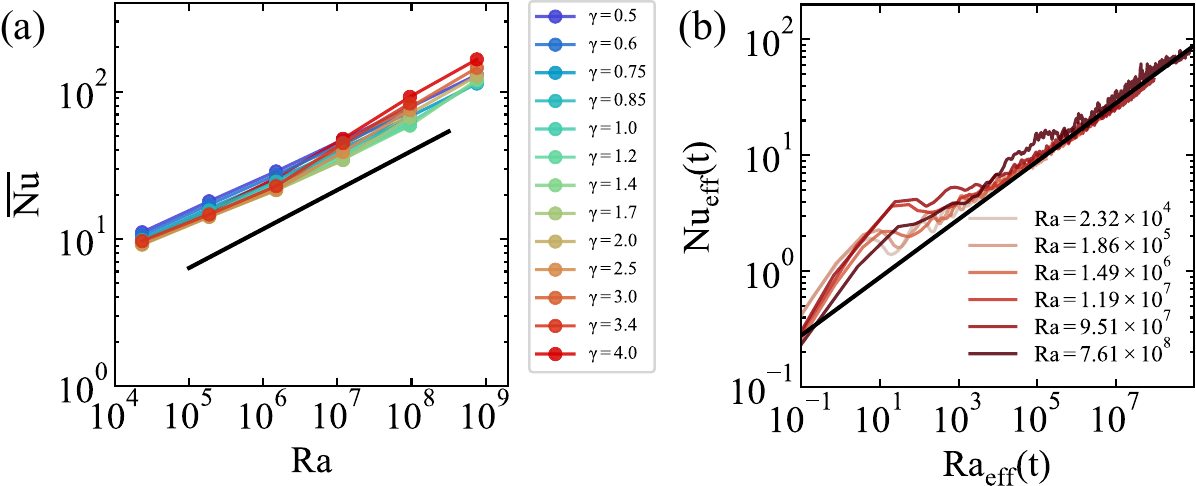}
  \caption{(a) The overall averaged Nusselt number $\overline{\text{Nu}}$ as a function of $\text{Ra}$ for different $\gamma$. (b) The effective Nusselt number $\text{Nu}_{\text{eff}}(t)$ as a function of the effective Rayleigh number $\text{Ra}_{\text{eff}}(t)$ (defined by the instantaneous size of ice) with different $\text{Ra}$ for $\gamma=1$. The lines are fits with exponent $0.265$.}
  \label{fig:suppfig2}
\end{figure}

Closely related to the Rayleigh number is the Nusselt number $\text{Nu}$. The Nusselt number is given by the ratio between convective and conductive heat transfer \cite{bejan1993}:
\begin{equation}
    \text{Nu}=\frac{hD}{k},
\end{equation}
where $h$ is the heat transfer coefficient and $k$ is the thermal conductivity. In the melting process, the melt rate is linked to $\text{Nu}$ through the Stefan boundary condition:
\begin{equation}
\tilde{v}_n=\frac{v_n}{U_0}=-\frac{1}{U_0}\frac{\kappa c_p}{\mathcal{L}}\frac{\partial T}{\partial n}=\frac{\overline{\text{Nu}}}{\text{Ste}\text{Ra}^{1/2}\text{Pr}^{1/2}A^{1/2}}.
\label{eq:vn}
\end{equation}
Here we estimate the averaged Nusselt number $\overline{\text{Nu}}$ using the overall melt rate (from initial to completely melted), and check how $\overline{\text{Nu}}$ is expressed as a function of $\text{Ra}$ using a scaling relation. In fig.~\ref{fig:suppfig2}(a), we plot the $\overline{\text{Nu}}$ as a function of $\text{Ra}$ for different aspect ratios, which all show a similar scaling as $\overline{\text{Nu}}\propto \text{Ra}^{0.265}$ by fitting. Fig.~\ref{fig:suppfig2}(b) shows the scaling between effective Rayleigh number $\text{Ra}_{\text{eff}}$ and effective Nusselt number $\text{Nu}_{\text{eff}}$ (both defined by the instantaneous size of the ice during melting), which also shows the same scaling as in (a).

\section*{Ice shape evolution}
\begin{figure}
\centering
  \includegraphics[width=0.8\columnwidth]{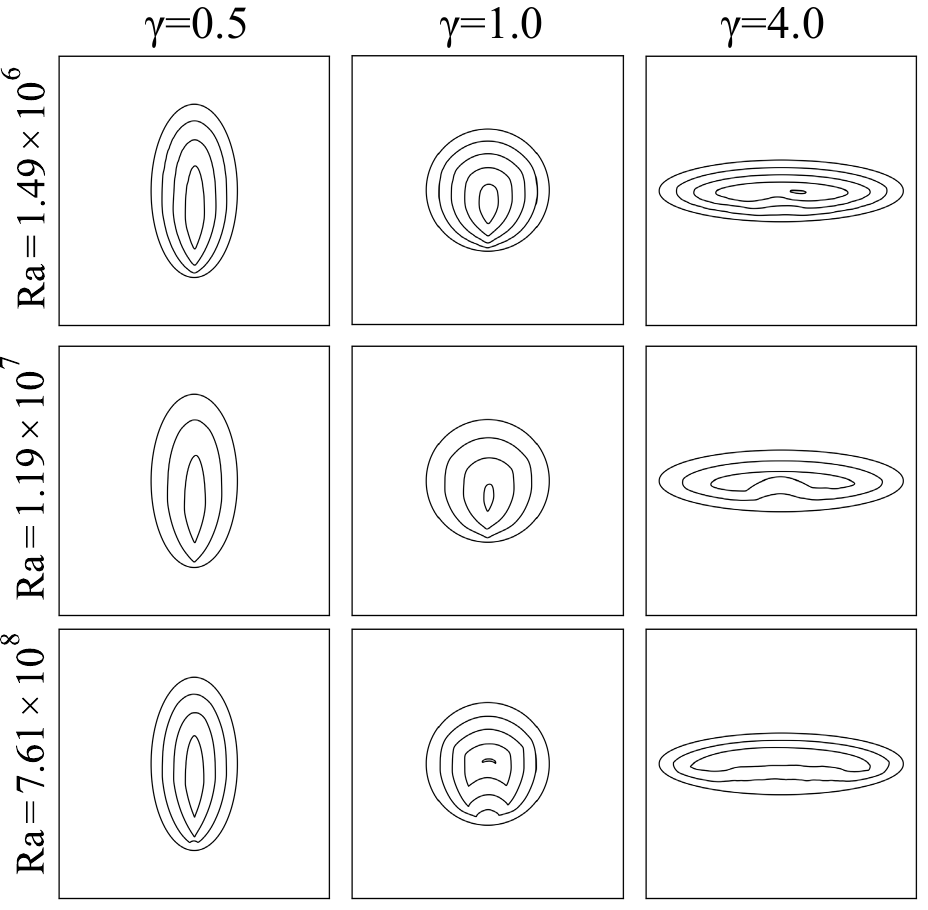}
  \caption{The contour evolution plot of ice for different starting $\gamma$ and $\text{Ra}$. The contours are taken equidistant in time.}
  \label{fig:suppfig3}
\end{figure}
\begin{figure}
\centering
  \includegraphics[width=0.75\columnwidth]{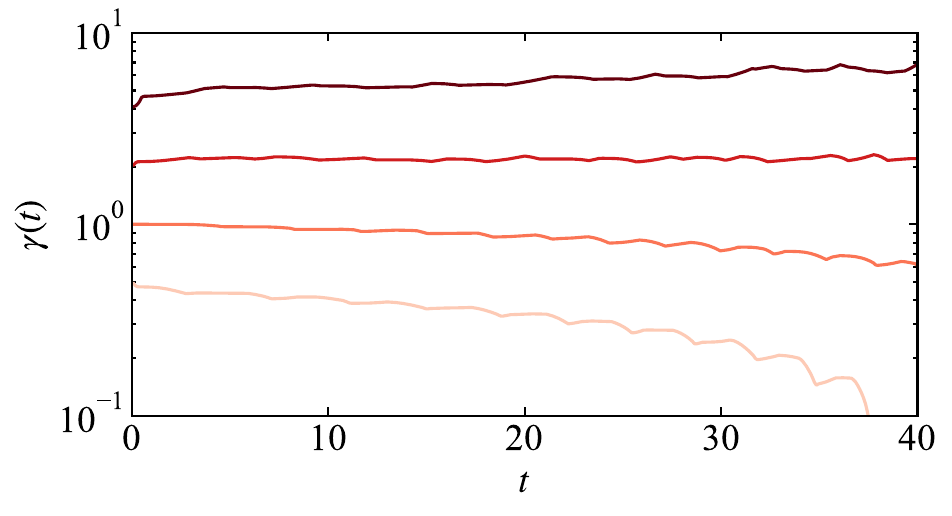}
  \caption{The evolution of the instantaneous aspect ratio $\gamma(t)$ of ice shape for different initial $\gamma=0.5,1,2,4$ for $\text{Ra}=1.86\times10^5$. $\gamma(t)$ is measured by the ratio of the maximum ice dimensions in vertical and horizontal directions. }
  \label{fig:suppfig4}
\end{figure}

Fig.~\ref{fig:suppfig3} shows the temporal evolution of the contours of ice shape for different initial $\gamma$ and $\text{Ra}$. One can see the cavity formation at high $\gamma$ and $\text{Ra}$ regime. Fig.~\ref{fig:suppfig4} shows the evolution of the aspect ratio of ice shape for different initial $\gamma$ at $\text{Ra}=1.86\times10^5$. One can see $\gamma(t)$ is actually changing with time, which explained the difference between our numerical results and the theoretical model (which assumes a constant $\gamma \neq \gamma(t)$).


\end{document}